\documentclass[apj]{emulateapj}

\usepackage{graphicx} 
\usepackage{epstopdf}

\begin{document}
\title{Discovery of a very high energy gamma-ray signal from the 3C~66A/B region}

%\input{magic_members_ApJ}
% authors 13.10.2008  Format ApJ
%
\author{
E.~Aliu\altaffilmark{a},
H.~Anderhub\altaffilmark{b},
L.~A.~Antonelli\altaffilmark{c},
P.~Antoranz\altaffilmark{d},
M.~Backes\altaffilmark{e},
C.~Baixeras\altaffilmark{f},
S.~Balestra\altaffilmark{d},
J.~A.~Barrio\altaffilmark{d},
H.~Bartko\altaffilmark{g},
D.~Bastieri\altaffilmark{h}
% ...
}
\altaffiltext{a} {IFAE, Edifici Cn., Campus UAB, E-08193 Bellaterra, Spain}
\altaffiltext{b} {ETH Zurich, CH-8093 Switzerland}
\altaffiltext{c} {INAF National Institute for Astrophysics, I-00136 Rome, Italy}
\altaffiltext{d} {Universidad Complutense, E-28040 Madrid, Spain}
\altaffiltext{e} {Technische Universit\"at Dortmund, D-44221 Dortmund, Germany}
\altaffiltext{f} {Universitat Aut\`onoma de Barcelona, E-08193 Bellaterra, Spain}
\altaffiltext{g} {Max-Planck-Institut f\"ur Physik, D-80805 M\"unchen, Germany}
\altaffiltext{h} {Universit\`a di Padova and INFN, I-35131 Padova, Italy}
% ...

%% abstract %%%%%%%%%%%%%%%%%%%%%%%%%%%%%%%%%%%%%%%%%%%%%%%%%

\begin{abstract}

The MAGIC telescope observed the region around the distant blazar 3C~66A for 54.2\,hr in 2007 August--December. 
The observations resulted in the discovery of a $\gamma$-ray source 
centered at celestial coordinates R.A. = $2^{\mathrm{h}} 23^{\mathrm{m}} 12^{\mathrm{s}}$ and decl.$=43^{\circ} 0.'7$ (MAGIC~J0223+430), coinciding with the nearby radio galaxy 3C~66B. 
A possible association of the excess with the blazar 3C~66A is discussed.
The energy spectrum of MAGIC~J0223+430 follows a power law with a normalization of $\left(1.7\pm 0.3_{\mathrm{stat}} \pm 0.6_{\mathrm{syst}} \right)\times10^{-11}$ 
TeV$^{-1}$ cm$^{-2}$ s$^{-1}$ at 300\,GeV and a photon index $\Gamma = -3.10 
\pm 0.31_{\mathrm{stat}} \pm 0.2_{\mathrm{syst}}$. 
\end{abstract}

\keywords{gamma rays: observations --- BL Lacertae objects: individual (3C~66A) --- galaxies: individual (3C~66B) --- ISM: individual (MAGIC~J0223+430)}

\section{Introduction}
\label{intro}

As of today, there are 23 known extragalactic very high energy (VHE, defined here as $E>100$\,GeV)
$\gamma$-ray sources. All of them are active galactic nuclei 
(AGNs) with relativistic jets. With the exception of the radio galaxy M~87 
all detected sources are blazars, whose jets (characterized by a bulk Lorentz factor $\Gamma \sim 20$) point, within a
small angle ($\theta \sim 1/\Gamma$), to the observer. The spectral energy
distribution (SED, logarithm of the observed energy density versus logarithm of the photon energy) 
of AGNs shows typically a two-bump structure. The lower-energy
bump originates from synchrotron radiation of relativistic electrons
spiraling in the magnetic field of the jet. For the origin of the high-frequency bump, 
various models have been
proposed, the most popular
invoking inverse Compton scattering of ambient photons. There have
been several suggestions for the origin of the low-frequency seed
photons that are up-scattered to $\gamma$-ray energies: they
may be produced within the jet by synchrotron radiation \citep[synchrotron self-Compton or SSC mechanism, e.g.][]{Maraschi,Bloom96} or come from outside the jet \citep[external Compton or EC mechanism, e.g.][]{Dermer}. Relativistic effects boost the observed emission 
as the Doppler factor depends on the angle to the line of sight.  
For sources with a large angle between the jet and the line of 
sight (e.g., the radio galaxy M~87), these classic inverse Compton 
scenarios cannot account for the VHE $\gamma$-ray emission.
In this case, models that depend less critically on beaming effects are needed 
\citep[e.g.][]{Neronov, Tavecchio}. The VHE $\gamma$-ray emission
of AGNs might also be of hadronic origin through the emission from secondary electrons \citep[e.g.][]{Mannheim, Mucke}. 

3C~66A and 3C~66B are two AGNs separated by just $6'$ in the sky.
3C~66B is a large
Fanaroff--Riley-I-type (FRI)
%\citep[FRI,][]{FR} 
radio galaxy, similar to M~87,
with a redshift of 0.0215 \citep{Stull}, whereas 3C~66A is a blazar
with uncertain redshift. The often referred redshift of 0.444
\citep{Miller} for 3C~66A is based on a single measurement of one emission
line only (and the authors were not certain on the realness of the
feature), while in later observations no lines in the
spectra of 3C~66A were reported \citep{Finke}. Based on the marginally resolved host
galaxy \citep{Wurtz}, a photometric redshift of $\sim 0.321$ was inferred.

3C~66A, a promising candidate for VHE $\gamma$-ray
emission, was observed several times with
satellite-borne and ground-based $\gamma$-ray detectors. The EGRET
source 3EG~J0222+4253 was associated with 3C~66A \citep{hartman}, but the
association was ambiguous because the error box is large enough to cover 3C~66B and the
nearby pulsar PSR~J0218+4232 \citep{Verbunt, Kuiper}. In 
the TeV regime the Crimean Astrophysical Observatory's GT-48 imaging
atmospheric Cerenkov telescope has claimed repeated detections of this source above
900\,GeV \citep{Neshpor, Stepanyan} with a flux as high as 
$\left(3\pm 1\right) \times 10^{-11}$\,cm$^{-2}$\,s$^{-1}$. HEGRA and
WHIPPLE reported upper limits, $F\left(>630\,\mathrm{GeV}\right)<1.42\times 10^{-11}$\,cm$^{-2}$\,s$^{-1}$ 
\citep{hegra} and  $F\left(>350\,\mathrm{GeV}\right)<0.59\times 10^{-11}$\,cm$^{-2}$\,s$^{-1}$ \citep{whipple}, 
from non-simultaneous observations. The STACEE solar array also provided an upper limit of $F\left(>184\,\mathrm{GeV}\right)<1.2 \times 10^{-10}$\,cm$^{-2}$\,s$^{-1}$ \citep{bramel}. In 2008 September, the Veritas collaboration reported a clear detection of 3C~66A~\citep{Atel1753} above 100\,GeV with an integral flux on the level of 10\% of the Crab Nebula flux.
Shortly after, a high state of 3C~66A was also reported by the Fermi Gamma-ray Space Telescope at energies above 20\,MeV \citep{fermiAtel}.

In this paper we report the discovery of VHE $\gamma$-ray emission located $6.'1$ away from the blazar 
3C~66A and coinciding with the radio galaxy 3C~66B in 2007. 
In Section 2, we describe the observations and the data analysis chain.
The results of the analysis are presented in Section 3 and discussed in Section 4.

%%%%%%%%%%%%%%%%%%%%%%%%%%%%%%%%%%%%%%%%%%%%%%%%%
%%%%%%%       OBSERVATIONS AND ANALYSIS                   %%%%%%%%%%%%%%%%%%
%%%%%%%%%%%%%%%%%%%%%%%%%%%%%%%%%%%%%%%%%%%%%%%%%

\section{Observations and Data Analysis}
\label{analysis}

3C~66A underwent an optical outburst in 2007 August, as monitored by the Tuorla 
blazar monitoring program. The outburst triggered VHE $\gamma$-ray 
observations of the source with the MAGIC telescope following the Target of Opportunity program,
which resulted in discoveries of new VHE $\gamma$-ray sources in the past
\citep{Albert06, Albert07a, magicS5}.

MAGIC has a standard trigger threshold of 60\,GeV, an angular
resolution of $ \sim 0.^\circ 1$ and an energy resolution
above 150\,GeV of $\sim 25\%$ (see \citet{crab} for details).

Data were taken in the false-source tracking (wobble) mode
\citep{Fomin1994} pointing alternatively to two different sky
directions, each at $24'$ distance from the 3C~66A catalog
position. The zenith distance distribution of the data extends from 13$^{\circ}$ to 35$^{\circ}$. Observations were made in 2007 August, September, and December and lasted 54.2\,hr, out of which 45.3\,hr passed the
quality cuts based on the event rate after image cleaning. 
An additional cut removed the events with total charge 
less than 150 photoelectrons (phe) 
in order to assure a better background rejection.

Just before the start of the observation campaign $\sim 5 \%$ of the mirrors on the telescope were replaced, worsening the optical point-spread function (PSF). As a consequence, a new calibration of the mirror alignment system became necessary, which took place within the observation campaign and improved the PSF again. 
The sigma of the Gaussian PSF (40\% light containment) was measured to be $3.'0$ in 2007 August 12-14, $2.'6$ in 2007 August 15-26 and $2.'1$ in 2007 September and December. 
To take this into account, data were analyzed separately for each period and the results were combined at the end of the analysis chain. However, the realignment resulted in a mispointing, which was taken care of by a new pointing model \citep{bdw} applied offline using starguider information \citep{bretz}. Considering the additional uncertainty caused by the offline corrections, we estimate the systematic uncertainty of the pointing accuracy to be $2'$ on average. Note that in the case of an optimal pointing model the systematic uncertainty is below $2'$, being $1'$ on average \citep{bdw,crab}.

The data analysis consists of several steps. 
Initially, a standard calibration of the data~\citep{NIMA} is performed. In the next step, an image cleaning procedure is  applied
using the amplitude and timing information of the calibrated
signals. In particular, the arrival times of the photons in core
pixels ($>6$\,phe) are  required to be within a time
window of 4.5\,ns and for boundary pixels ($>3$\,phe) within a time
window of 1.5\,ns from a neighboring core pixel. 
For the surviving pixels of each event image parameters are calculated \citep{Hillas1985}.
Using the good time resolution of the recorded signals ($\sim 400$\,ps), unique to MAGIC, the time gradient along the main shower
axis and the time spread of the shower pixels are computed \citep{timing}. Hadronic background suppression is achieved
using the Random Forest (RF) method~\citep{Bock2007},
where for each event the so-called \textsc{Hadronness} parameter is
computed, based on the image and the time parameters. 
Moreover, the RF method
is used for the energy estimation trained on a Monte Carlo simulated
$\gamma$-ray sample with the same zenith angle distribution as the data
sample.

\begin{figure}[t]
\centering
\includegraphics*[width=1.\columnwidth]{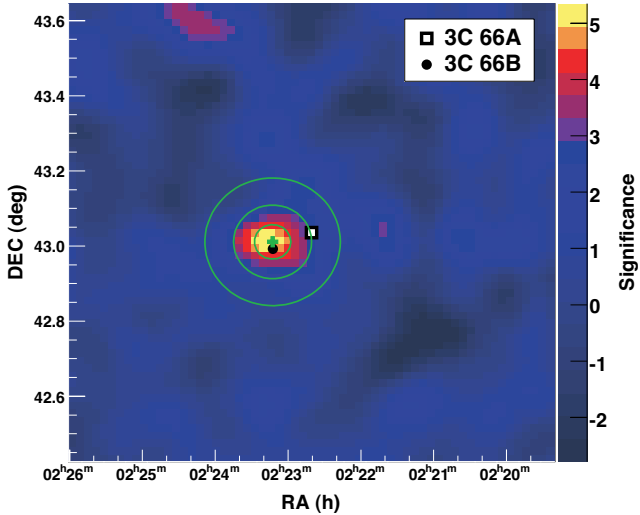}
%\plotone{f1.eps}
\caption{Significance map for $\gamma$-like events above 150\,GeV 
in the observed sky region. The green cross 
corresponds to the fitted maximum excess position of MAGIC~J0223+403. 
The probability of the true source to be inside 
the green circles is 68.2\%, 95.4\%, and 99.7\% for the inner,
middle, and outer contour, respectively. 
The catalog positions 
of 3C~66A and 3C~66B are indicated by a white square and a black dot, respectively.
}
\label{fig:skymap}

\end{figure}

%%%%%%%%%%%%%%%%%%%%%%%%%%%%%%%%%%%%%
%%%%%%%%%%%%%%%   RESULTS    %%%%%%%%%%%%%%%
%%%%%%%%%%%%%%%%%%%%%%%%%%%%%%%%%%%%%

\section{Results}
\label{results}

Figure \ref{fig:skymap} shows a significance map produced from the signal and background maps, 
both smoothed with a Gaussian of $\sigma=6'$ 
(corresponding to the $\gamma$-PSF), for photon energies between 150\,GeV and 1\,TeV. 
%% Added sentence
For the background rejection a loose cut in the \textsc{Hadronness} parameter is applied to keep a large number of gamma-like events.
The center of gravity of the $\gamma$-ray emission is derived from Figure~\ref{fig:skymap} by fitting a bell-shaped function of the form
\begin{equation}
F(x,y) = A \cdot \exp \left[ -\frac{(x-\bar{x})^2+(y-\bar{y})^2}{2\sigma^2}\right]
\label{gaus}
\end{equation}
for which the distribution of the excess events is assumed to be rotationally symmetric, i.e., $\sigma_x=\sigma_y=\sigma$. 
The fit yields reconstructed coordinates of the excess center of R.A. = $2^{\mathrm{h}} 23^{\mathrm{m}} 12^{\mathrm{s}}$ and decl.$=43^{\circ} 0.'7$.
The detected excess, which we name MAGIC~J0223+430, is $6.'1$ away from the catalog position of 3C~66A, 
while the distance to 3C~66B is $1.'1$. 

In order to estimate the statistical uncertainty of the reconstructed position, we simulated $10^4$ sky maps with the same number
of background and excess events as in the data. The excess position in the sky maps was fitted and the distance to the simulated source position calculated.
From the histogram of the distances we obtained probabilities for an offset between the true source and the fit to the excess.
The probabilities shown in Figure~\ref{fig:skymap} by the green contours 
correspond to 68.2\%, 95.4\%, and 99.7\% for the inner,
middle, and outer contour, respectively.
Using this study we find that the measured excess coincides with the catalog position of 3C~66B.
The origin of the emission from 3C~66A can be statistically excluded with a probability of 95.6\%.
Adding linearly the systematic uncertainty of the pointing of the data set ($2'$, see above), i.e.,\ shifting the excess position by $2'$ toward the catalog position of 3C~66A, the exclusion probability is 85.4\%.

\begin{figure}[t]
\centering
\includegraphics*[width=1.\columnwidth]{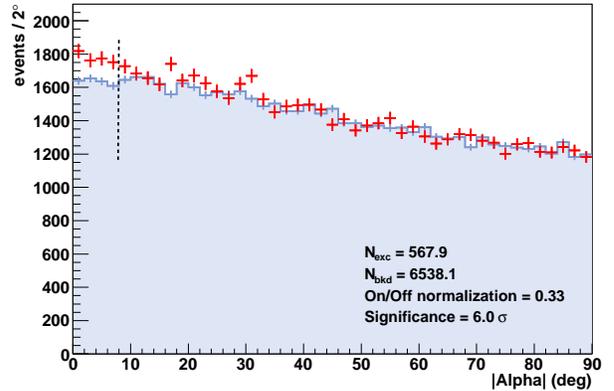}
%\plotone{f2.eps}
\caption{\textsc{$\mid$Alpha$\mid$} distribution after all cuts evaluated with respect to the position of MAGIC~J0223+430. A $\gamma$-ray excess with a significance of $6.0\,\sigma$ is found, which corresponds to a post-trial significance of $5.4\,\sigma$.}
\label{fig:alpha}
\end{figure}

To calculate the significance of the detection, 
an \textsc{$\mid$Alpha$\mid$} distribution was
produced, where \textsc{Alpha} is the angle between the major axis of the
shower image ellipse and the source position in the camera. For the calculation of the source-dependent image parameters we considered the fitted position of the excess.
Background rejection was
achieved by a cut in \textsc{Hadronness}, which was optimized
using Crab Nebula data taken in similar conditions and diluted to 5\% of its real flux. The cut in
\textsc{$\mid$Alpha$\mid$} that defines the signal region was also optimized in the same way. 
The \textsc{$\mid$Alpha$\mid$} and \textsc{Hadronness} cuts together have an efficiency of 40\% in keeping Monte Carlo simulated $\gamma$ events, and result in an energy threshold of approximately 230\,GeV.
%% Added footnote
\footnote[1]{Defined as the peak of the distribution of Monte Carlo generated gamma-ray events after all cuts.} 
A signal of $6.0\,\sigma$ significance (pre-trial) was found (see Figure~\ref{fig:alpha}).
We estimated the number of trials of the signal search by projecting the $\gamma$-ray acceptance of the camera into the field of view of the observations, and defined the search region where the $\gamma$-ray acceptance after cuts is larger than 50\%. In this way, we obtained an area of $2.18\, \mathrm{deg}^{2}$.
Given that the 68\% containment radius for $\gamma$-rays from a point-like source is $0.^{\circ}152$, we calculated the number of independent trials to be 30.
%Given that the $\gamma$-PSF of the analysis is $0.10^{\circ}$ (40\% containment radius) we calculated the number of independent trials to be 30.\footnote[2]{Defined as the area of the search region divided by the 68\% containment area of an individual point-like source.} Correcting for the number of trials we find a post-trial significance of the measured signal of $5.4\,\sigma$. 

\begin{figure}[t]
\begin{center}
\includegraphics*[width=1.\columnwidth]{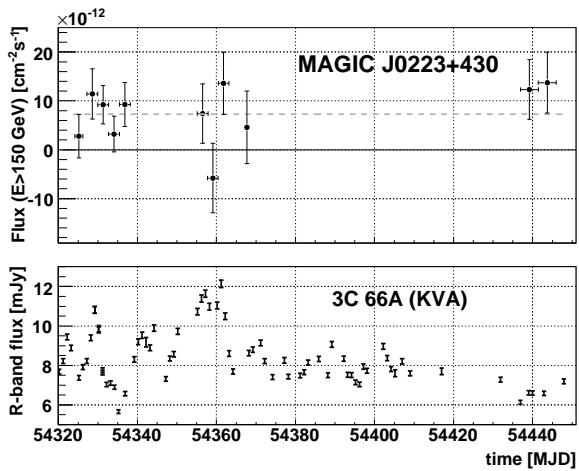}
%\plotone{f3.eps}
\caption{Light curve of MAGIC~J0223+430. Upper panel: MAGIC integral flux above 150\,GeV in bins of 3 days (except for periods where the sampling was coarser). The gray dashed line indicates the average $\gamma$-ray flux. Lower panel: optical light curve of 3C~66A as measured by the KVA telescope.
During the MAGIC observations 3C~66A was very bright at optical
wavelengths varying from 6\,mJy to 12\,mJy in the $R$-band (the baseline flux in the historical data being $\sim 6$\,mJy). 
In the same period the optical flux of 3C~66B remained constant, which is a typical behavior for large radio galaxies.}
\label{fig:lightcurve}
\end{center}
\end{figure}

Figure~\ref{fig:lightcurve} shows the light curve of MAGIC~J0223+430 together with the flux of 3C~66A in optical wavelengths. 
As we integrate over $\gamma$-ray events from a wide sky region ($\sim0.07\, \mathrm{deg}^2$), we cannot exclude that 3C~66A contributes to the measured signal. 
The integral flux above 150\,GeV corresponds to $\left(7.3\pm 1.5\right) \times 10^{-12}$\,cm$^{-2}$\,s$^{-1}$ (2.2\% of the Crab Nebula flux) and is the lowest ever detected by MAGIC. 
The $\gamma$-ray light curve is consistent with a constant flux within statistical errors. These errors, however, are large, and some variability of the signal cannot be excluded.
%{\bf The $\gamma$-ray light curve shows no statistically significant variability, but due to large errorbars firm conclusion about the variability of the signal is not possible.}
%No significant variability in the $\gamma$-ray light curve was observed during the MAGIC observations whereas 
%the blazar 3C~66A was in high optical state varying from 6\,mJy to 12\,mJy in the $R$-band (the baseline flux in the historical data being $\sim 6$\,mJy). 
%In the same period the optical flux of 3C~66B remained constant, which is a typical behavior for large radio galaxies.
%THE OPTICAL VARIABILITY I MOVED TO CAPTION

\begin{figure}[t]
\begin{center}
\includegraphics*[width=1.\columnwidth]{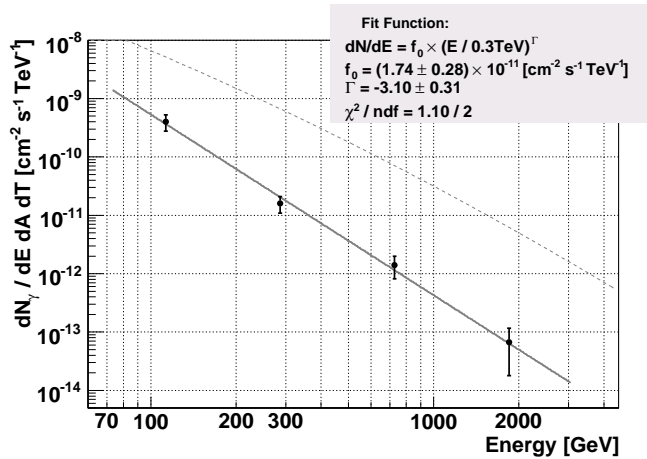}
%\plotone{f4.eps}
\caption{Differential energy spectrum of MAGIC~J0223+430. The fit to the data is shown by the solid gray line and the fit parameters are listed in the inset. No correction for the $\gamma-\gamma$ attenuation due to the EBL has been made. The Crab Nebula spectrum \citep{crab} is also shown as a reference (dashed gray line).}
\label{fig:spectrum}
\end{center}
\end{figure}

For the energy spectrum of MAGIC~J0223+430, loose cuts are made to keep the $\gamma$-ray acceptance high. The differential energy 
spectrum was unfolded using the Tikhonov unfolding technique \citep{tikhonov,unfolding} and is shown in Fig.~\ref{fig:spectrum}. The spectrum can be well fitted by a power law
which gives a differential flux (TeV$^{-1}$ cm$^{-2}$ s$^{-1}$) of:
\begin{equation}
\frac{\mathrm {d}N}{\mathrm{d}E\, \mathrm {d}A\, \mathrm {d}t} = (1.7\pm 0.3)\times10^{-11}(E/300\,\mathrm {GeV})^{-3.1\pm0.3} 
\end{equation}
The quoted errors are statistical only. The systematic uncertainty is
estimated to be 35\% in the flux level and 0.2 in the power law photon
index \citep{crab}.  
As in the case of the light curve, we cannot exclude that 3C~66A contributes to the measured signal. Thus, the spectrum shown in Figure~\ref{fig:spectrum} represents a combined 
$\gamma$-ray spectrum from the observed region.

%%%%%%%%%%%%%%%%%%%%%
% DISCUSSION
%%%%%%%%%%%%%%%%%%%%%

\section{Discussion and conclusions}
\label{conclusions}

A new VHE $\gamma$-ray source MAGIC~J0223+430 was detected in 2007 August to December.
Given the position of the excess measured by MAGIC above 150\,GeV, the source of the $\gamma$-rays is most likely 3C~66B. The VHE $\gamma$-ray flux was found to be on the level of 2.2\% Crab Nebula flux and was constant during the observations.
The differential spectrum of MAGIC~J0223+430 has a photon spectral index of $\Gamma=3.10\pm 0.31$ and extends up to $\sim 2$\,TeV. 
In view of the recent detection of 3C~66A at VHE $\gamma$-rays \citep{Atel1753}, we note
that if 3C~66A was emitting $\gamma$-rays in 2007 August to December then its flux 
was at a significantly lower level than in 2008. 
We also note that we cannot exclude the scenario suggested in a recent work by \citet{TavGhis} 
that the observed spectrum would be a combination of emission from 3C~66B (dominating at energies above 150\,GeV) 
and blazar 3C~66A (at lower energies).

In the unlikely case, excluded with probability 85.4\%, that the total signal
and observed spectrum presented in this paper originates from 3C~66A,
the redshift of the source is likely to be significantly lower than
previously assumed. 
Due to the energy-dependent absorption of VHE $\gamma$-rays with low-energy photons of the extragalactic background
\citep[EBL,][]{gould}, the VHE $\gamma$-ray flux of distant sources is significantly suppressed.
We investigated the measured spectrum by MAGIC following the prescription
of \citet{mazin}, and derived a redshift upper limit of the source to be $z<0.17$ ($z<0.24$) under the 
assumption that the intrinsic energy spectrum cannot be
harder than $\Gamma = 1.5$ ( $\Gamma = 0.666$).
%Indeed, due to the energy-dependent absorption of VHE
%$\gamma$-rays with low energy photons of the extragalactic background
%light \citep[EBL,][]{gould}, more than 99\% of $\gamma$-rays with
%energies above 1 TeV would be absorbed if $z>0.23$. This derived redshift limit is valid under
%the assumption that the intrinsic VHE $\gamma$-ray flux cannot be
%harder than $\Gamma = 1.5$ \citep[see][]{mazin}.
This assumption of $\Gamma > 1.5$ is based on particle acceleration
arguments \citep{AharonianEBL}, and the fact that none of the sources
in the EGRET energy band (not affected by the EBL) have shown a harder
spectrum. The latter assumption of $\Gamma > 0.666$ can be considered as an extreme case of the spectrum hardness, 
suggesting a monochromatic spectrum of electrons when interacting with a soft photon target field \citep{katarzynski}.\footnote[2]{ 
See also \citet{stecker} for more detailed calculations.}
If $z>0.24$ for 3C~66A,
an alternative explanation for a hard
intrinsic spectrum at energies above 100\,GeV can be given if
$\gamma$-rays are passing through a narrow band of optical-infrared
photons in the vicinity of the blazar. Such narrow radiation fields
can produce arbitrarily hard intrinsic spectra by absorbing specific
energies of $\gamma$-rays \citep{aha_abs}. We also note that, in this
case, the intrinsic VHE luminosity of 3C~66A should exceed
$10^{47}$\,erg\,s$^{-1}$, which is an unusually large value for a BL~Lac
object \citep{robert}, also in view of its spectral characteristics \citep{persic}.

\newcommand{\degree}{{}^{\circ}}

3C~66B is a FRI radio galaxy similar to M~87, which has been detected to
emit VHE $\gamma$-rays \citep{hegraM87,m87}. 
Since the distance of 3C~66B is 85.5\,Mpc, its intrinsic VHE luminosity would be two to eight times
higher than the one of M~87 (22.5\,Mpc) given the reported variability of M~87 \citep{m87,MAGICm87}.

As in the case of M~87, there would be several possibilities for the region responsible of 
the TeV radiation in 3C~66B: the vicinity
of the supermassive black hole \citep{Neronov}, the unresolved base of the
jet \citep[in analogy with blazar emission models;][]{Tavecchio} 
and the resolved jet. Unlike for M~87, we do not observe
significant variability in the VHE $\gamma$-ray flux and therefore we have
no constrains on the size of the emission region. However, as the
angle to line of sight is even larger than in M~87 (M~87: 19$\degree$,
\citet{Perlman}; 3C~66B: 45$\degree$, \citet{giovanni}) the resolved jet
seems an unlikely site of the emission. On the other hand, the unresolved base of the 
jet seems a likely candidate for the emission site as 
it could point with a smaller angle 
to the line of sight. 
If the viewing angle was small, blazar-like emission mechanisms cannot be
excluded. The orbital motion of 3C~66B shows evidence for a
supermassive black hole binary (SMBHB) with a period of
$1.05\pm0.03$ years \citep{sudou}. The SMBHB would likely cause the   
jet to be helical, and the pointing direction of the unresolved jet could
differ significantly from the direction of the resolved jet.

Given the likely association of MAGIC~J0223+430 with 3C~66B, our detection would 
establish radio galaxies as a new class of VHE $\gamma$-ray emitting sources.
According to \cite{Ghisellini05}, there are eight FR I radio galaxies in the 3CR 
catalog that should have a higher $\gamma$-ray flux at 100\,MeV than 3C 66B,
but possibly many of these 
sources are rather weak in the VHE $\gamma$-ray band. Further observations 
of radio galaxies with the Fermi Gamma-ray Space Telescope as well as by ground-based telescopes are needed 
to further study the $\gamma$-ray emission properties of radio galaxies.

\vspace{0.1cm}

We thank the Instituto de Astrofisica de Canarias for the excellent working conditions at the 
Observatorio del Roque de los Muchachos in La Palma. The support of the German 
BMBF and MPG, the Italian INFN, and Spanish MCINN is gratefully
acknowledged. This work was also supported by ETH Research Grant 
TH 34/043, by the Polish MNiSzW Grant N N203 390834, and by the YIP of the 
Helmholtz Gemeinschaft.

\end{document}